\documentclass[preprint,proceedings]{rmaa}

\SetYear{2003}
\SetConfTitle{Compact Binaries in the Galaxy and Beyond}

\title{SDSS J1240$-$01: A New AM CVn Candidate from the Sloan Digital Sky Survey}

\author{G. H. A. Roelofs,\altaffilmark{1} P. J. Groot,\altaffilmark{1} D. Steeghs,\altaffilmark{2} and G. Nelemans\altaffilmark{3}}

\altaffiltext{1}{Dep.\ of Astrophysics, Univ.\ of Nijmegen, P.O. Box 9010, 6500~GL Nijmegen, The Netherlands (\protect\email{gijsroel@astro.kun.nl}).}
\altaffiltext{2}{Harvard-Smithsonian CfA, Cambridge, USA}
\altaffiltext{3}{Institute of Astronomy, Cambridge, UK}

\suppressfulladdresses

\listofauthors{G. H. A. Roelofs, P. J. Groot, D. Steeghs, \& G. Nelemans}
\indexauthor{Roelofs, G. H. A.}
\indexauthor{Groot, P. J.}
\indexauthor{Steeghs, D.}
\indexauthor{Nelemans, G.}

\addkeyword{Stars: binaries: close}
\addkeyword{Stars: individual: AM Canum Venaticorum}
\addkeyword{Stars: individual: SDSS J124058.03-015919.2}

\begin{document}
\maketitle 

\boldabstract{A better understanding of the AM CVn population is crucial to constrain their candidacy as SN Ia progenitors, to test binary evolution (in particular the common-envelope phase), and to predict their observable gravitational radiation signature. An AM CVn-dedicated search in the Sloan Digital Sky Survey-DR1 resulted in the discovery of SDSS J124058.03$-$015919.2, a new AM CVn candidate previously identified as a DB white dwarf in the 2dF quasar survey.}

Both the SDSS and 2dF spectra show double-peaked helium emission lines and absence of any such hydrogen lines, indicating a helium-dominated accretion disk. They further show broad absorption features in the blue, which resulted in its DB white dwarf classification. The continuum can be fitted well with a 17,000\,K blackbody. The system appears to be quite old and reminds of GP Com and CE 315 (low mass transfer; optically thin disk) but its still quite hot white dwarf primary, possibly re-heated by recent high mass transfer, indicates a much younger system.

Our first optical follow-up (taken 13-12-2003 with Magellan-I, spectral resolution 3\,\AA) clearly shows the double-peaked \ion{He}{i} emission lines as well as \ion{He}{ii}\,$\lambda$4686 and \ion{N}{iii}\,$\lambda$4634$+$4640. This combination suggests either a Bowen fluorescence mechanism at work (cf.\ Casares et al.\ 2003) or an extreme nitrogen abundance in the system (cf.\ G\"ansicke et al.\ 2003). \ion{He}{ii}\,$\lambda$4686 is observed in most AM CVns, although in SDSS J1240 it is unusually strong compared to \ion{He}{i}\,$\lambda$4713 (equivalent widths $-4.0$\,\AA\ and $-4.4$\,\AA\ respectively, $\pm$10\%). No traces of \ion{N}{iii}\,$\lambda$4634$+$4640 are found in high-quality spectra of GP Com and CE 315, while in the new system it is remarkably strong at an equivalent width of $-2.4$\,\AA, more than half that of the \ion{He}{i}\,$\lambda$4713 line. The FWHM of these lines is only a quarter that of the helium lines ($\sim$5\,\AA\ versus $\sim$20\,\AA), which suggests a non-disk origin. The strongest helium line at 5875\,\AA\ has an equivalent width of $-31$\,\AA\ compared to $-78$\,\AA\ in GP Com, which can be explained nicely with an equally luminous, optically thin disk contributing little to the continuum, plus a primary that is more luminous by the factor expected from its higher temperature (17,000\,K versus 11,000\,K for GP Com, Marsh et al.\ 1991).

The AM CVn nature of the new system has yet to be proved beyond doubt with spectroscopic follow-up giving its orbital period, which we expect to be between 30--40 minutes. This places it between the cooler, shortest-period emission-line system GP Com and the longest-period systems among the outbursting AM CVns.

\begin{figure}[!ht]
\begin{center}
  \includegraphics[angle=270,width=\columnwidth]{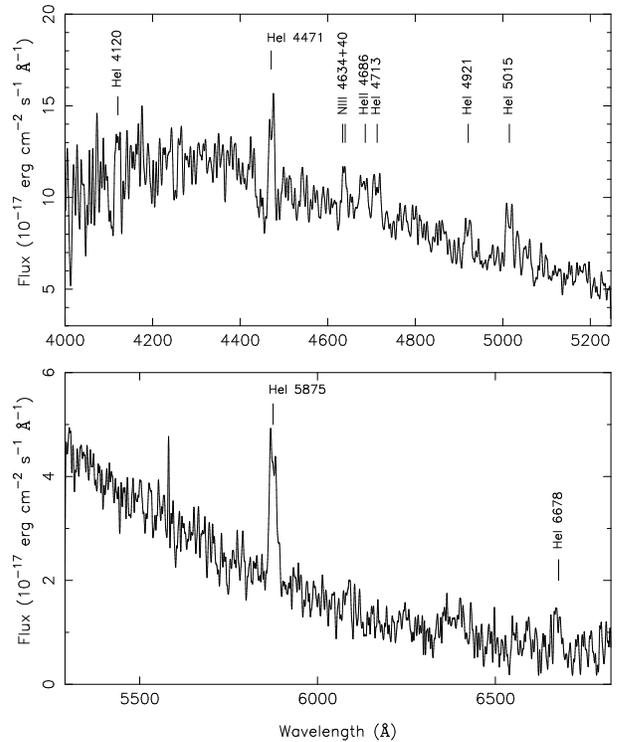}
  \caption{The new AM CVn candidate SDSS J1240$-$01, observed 13-12-2003 with IMACS at Magellan-I.}
  \label{groelofs_fig1}
\end{center}
\end{figure}

\end{document}